\documentclass[journal]{IEEEtran}
\usepackage{amsmath,amssymb}
\usepackage{algorithm}
\usepackage{algpseudocode}
\usepackage{cite}
\usepackage{graphicx}
\usepackage{placeins}
\usepackage{tikz} 
\usetikzlibrary{calc}
\usepackage{hyperref}
\usepackage{cleveref}
\crefname{figure}{Fig.}{Figs.}    
\crefname{subfigure}{Fig.}{Figs.} 
\creflabelformat{subfigure}{#2#1#3} 
\usepackage{booktabs}
\usepackage{microtype}
\usepackage{setspace}
\usepackage{xcolor}
\usepackage{algpseudocode}
\usepackage{subcaption}
\allowdisplaybreaks
\begin{document}
	
	\title{Rotatable Antenna Meets Multiple Access: \\NOMA or OMA?}
	
		\author{
		\IEEEauthorblockN{Qi Dai, Beixiong Zheng\textit{, Senior Member, IEEE}, Yanhua Tan, Weidong Mei\textit{, Member, IEEE},\\ Jie Tang\textit{, Senior Member, IEEE}, and Fangjiong Chen\textit{, Member, IEEE}}
		\thanks{Qi Dai, Beixiong Zheng, and Yanhua Tan are with the School of Microelectronics, South China University of Technology, Guangzhou 511442, China (E-mails: 7qidai@gmail.com; bxzheng@scut.edu.cn; tanyanhua06@163.com).}
        \thanks{Weidong Mei is with the National Key Laboratory of Wireless Communications, University of Electronic Science and Technology of China, Chengdu, 611731, China (E-mail: wmei@uestc.edu.cn).}
        \thanks{Jie Tang and Fangjiong Chen are with School of Electronic and Information Engineering, South China University of Technology, Guangzhou, 510640 China (E-mails: eejtang@scut.edu.cn; eefjchen@scut.edu.cn).}
	}
	
	\maketitle
	
	\begin{abstract}
		Rotatable antenna (RA) technology has emerged as a promising solution to enhance spectrum efficiency by exploiting additional spatial degrees of freedom (DoFs) in multiple access networks. However, the relative performance superiority among different multiple access schemes remains largely unclear due to the unique capability of RA in reconfiguring the directional gain pattern. In this letter, we conduct a theoretical comparison between non-orthogonal multiple access (NOMA) and orthogonal multiple access (OMA) schemes in RA-assisted communication systems in terms of transmit power minimization, subject to constraints on antenna rotational range and users’ target rates. To address the associated non-convex optimization problem, a particle swarm optimization (PSO) algorithm is employed to optimize the rotational angle. Simulation results demonstrate that RA-assisted schemes significantly reduce transmit power compared to fixed-antenna benchmarks. Furthermore, RA-assisted NOMA may perform worse than time-division multiple access (TDMA) for symmetric user deployments, while it exhibits superior robustness and energy efficiency in asymmetric scenarios.
	\end{abstract}
	
	
	\section{Introduction}
	The evolution toward next-generation wireless communication systems is driven by the explosive growth of mobile devices and the stringent requirements for ultra-high capacity, massive connectivity, and energy efficiency \cite{6GMIMO}. To meet these demands, non-orthogonal multiple access (NOMA) has been recognized as a key enabler for improving spectral efficiency (SE). Unlike conventional orthogonal multiple access (OMA) schemes, such as time-division multiple access (TDMA) and frequency-division multiple access (FDMA) \cite{OMA}, which avoids inter-user interference by assigning distinct users into orthogonal time or frequency resource blocks, NOMA allows multiple users to share the same time-frequency resource block by employing superposition coding (SC) at the transmitter and successive interference cancellation (SIC) at the receiver \cite{DaiNOMA}. To further enhance NOMA performance, numerous advanced technologies have been explored, such as unmanned aerial vehicles (UAV) \cite{UAVNOMA} and intelligent reflecting surfaces (IRS)\cite{IRSzheng}. However, the performance of these multiple-access schemes is intrinsically constrained by the static characteristic of fixed antennas (FAs), which limits the spatial degrees of freedom (DoFs). As a result, NOMA with FAs cannot dynamically adapt to user channel conditions, thus hindering the effective construction of channel gain disparities necessary for efficient SIC. Furthermore, enhancing channel quality by scaling up fixed antenna arrays leads to substantial increases in hardware cost, circuit power consumption, and computational complexity.
    
		\begin{figure}[!t] 
		\centering
		\includegraphics[width=3.0in]{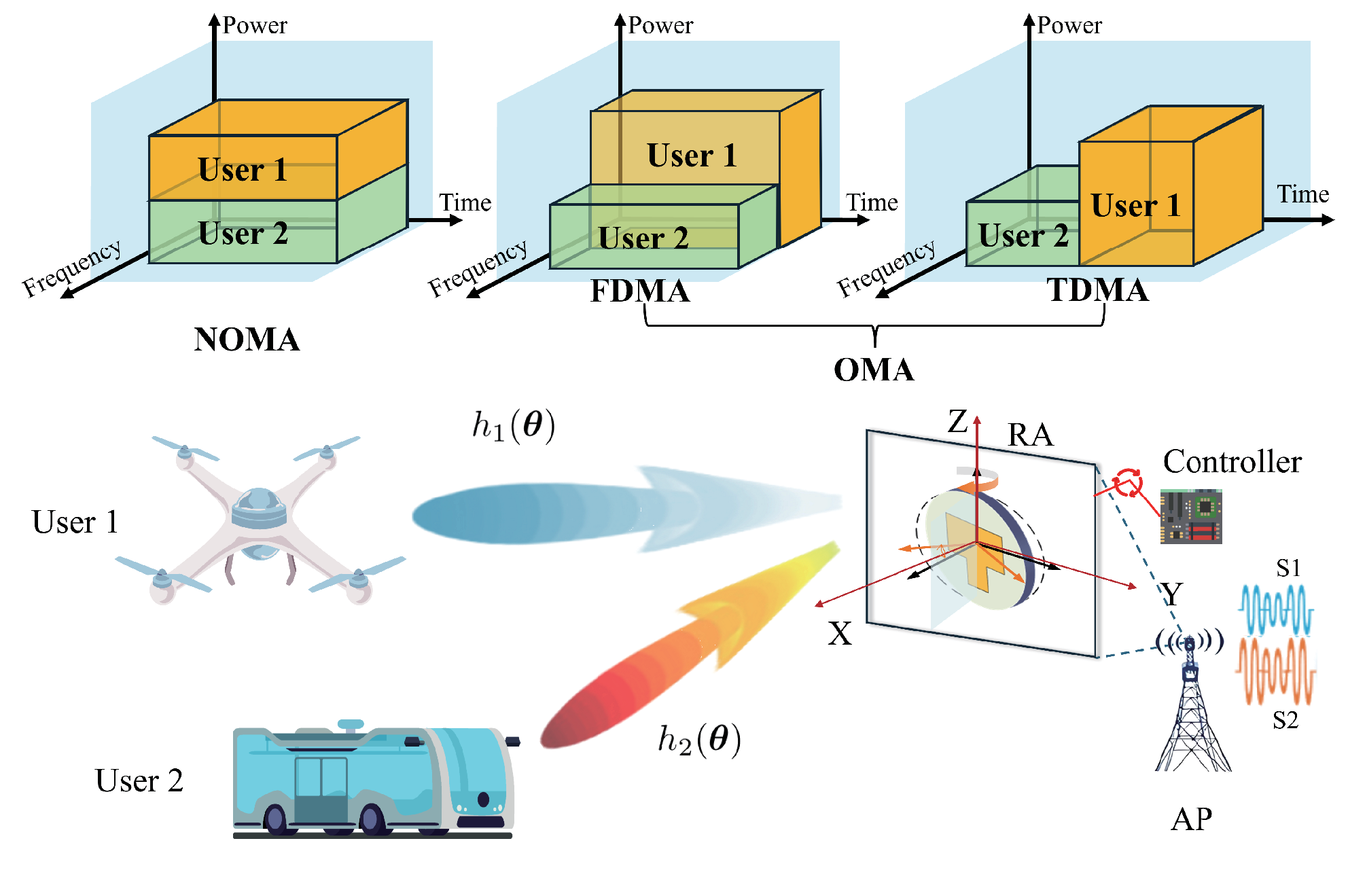} 
		\caption{An illustration of NOMA and OMA in an RA-assisted communication system with two users.}
		\label{fig:system}
	\end{figure}	
	To overcome the above-mentioned limitations, rotatable antenna (RA) has recently emerged as a revolutionary technology, garnering significant attention for its ability to dynamically adjust the boresight direction of each antenna to exploit additional spatial DoFs\cite{RAoverview2025}. Particularly, RA retains only the rotational capability to offer a highly cost-effective and compact solution that facilitates versatile beam steering and spatial reconfigurability. Specifically, by mechanically or electronically adjusting the three-dimensional (3D) orientation/boresight of individual antennas, RA arrays can dynamically reconfigure the directional gain pattern to actively enhance the array gains in desired directions, thereby significantly improving channel conditions. Owing to these attractive advantages, a growing body of research has been dedicated to establishing the theoretical and practical foundations of RA~\cite{zheng2026tutorial,zheng2025rotatableantennaenabledwireless,wumodeling,RAchannelestimation,dai2025rotatableantennaenabledsecurewireless,DaiCovert,zhou2025rotatableantennasintegratedsensing,tyhCR}. For instance, initial foundational frameworks encompassing system modeling, channel characterization, and performance analysis for RA-enabled communications were established in \cite{zheng2025rotatableantennaenabledwireless} and \cite{wumodeling}. Building upon these cornerstones, a comprehensive tutorial on RA technology was provided in \cite{zheng2026tutorial}, thoroughly illustrating its fundamentals, architectures, and applications. To facilitate practical implementation, a tailored channel estimation scheme has also been developed to significantly enhance estimation accuracy \cite{RAchannelestimation}. Moreover, studies on RA have been extended to a wide range of areas, including secure and covert wireless communications \cite{dai2025rotatableantennaenabledsecurewireless,DaiCovert}, integrated sensing and communication (ISAC) \cite{zhou2025rotatableantennasintegratedsensing}, and spectrum sharing in cognitive radio (CR) systems\cite{tyhCR}. Despite the above advances, the integration of RA into NOMA/OMA networks remains largely unexplored. In particular, a straightforward theoretical performance comparison between NOMA and OMA in RA-assisted systems is still lacking and inherently non-trivial, as the dynamic coupling between antenna orientation and multi-user resource allocation makes the energy-efficiency hierarchy between these schemes less predictable.

	Motivated by the above, this letter theoretically compares the performance of NOMA and OMA in an RA-assisted downlink communication system, where two types of OMA schemes are considered, i.e., FDMA and TDMA, as shown in \cref{fig:system}. To this end, we formulate the transmit power minimization problems for these schemes, subject to the users' target rate requirements and the practical rotational range constraints of RAs. By deriving the closed-form expressions for the minimum transmit power, we prove that RA-assisted NOMA and TDMA systems always outperform the RA-assisted FDMA system. Moreover, to tackle the non-convex optimization problem, a particle swarm optimization (PSO)-based algorithm is employed to obtain high-quality suboptimal solutions. Numerical results consolidate our theoretical findings, further verifying that the proposed RA technology can achieve a significant reduction in transmit power consumption.

	\section{System Model and Problem Formulation} 
	\label{sec:system}
	As illustrated in \cref{fig:system}, we consider the downlink of an RA-assisted communication system in a vehicular network environment, where a roadside access point (AP) equipped with a single RA serves multiple single-antenna users, such as unmanned aerial vehicles (UAVs) and autonomous ground vehicles, by employing either NOMA or conventional OMA schemes (i.e., FDMA and TDMA).\footnote{While we focus on a fundamental two-user scenario to tractably reveal the interplay between antenna rotation and different multiple access schemes, the analysis provides essential insights that can be readily extended to general multi-user cases via the user pairing concept.} To ensure a fair comparison, FDMA and TDMA serve the two users over orthogonal resource blocks (RBs) separated in the frequency and time domains, respectively, whereas NOMA serves both users simultaneously utilizing the entire time-frequency block. Without loss of generality, we assume that the AP is located at the origin of the 3D Cartesian coordinate system. Let $r_k = {\left\|\mathbf{u}_{k}\right\|}$ represent the distance between user $k$ and the AP, with $\mathbf{u}_k \in \mathbb{R}^{3 \times 1}, k \in \{1,2\}$ denoting the position of user $k$. Consequently, the unit-norm direction vector between the RA and user $k$ is given by \(\mathbf{\tilde{u}}_{k} = {\mathbf{u}_{k}}/{r_k}\).
	
	\subsection{Antenna Boresight Rotation Model}
	To mathematically characterize the rotational capability, the 3D boresight direction of the RA can be represented by a pointing vector, defined as
		\small	\begin{equation}
		\vec{\mathbf{f}} = [f_{x}, f_{y}, f_{z}]^T \in \mathbb{R}^{3 \times 1},
	\end{equation} \normalsize
	where \(f_{x}\), \(f_{y}\), and \(f_{z}\) denote the projections of the RA's pointing vector on the $x$-, $y$-, and $z$-axes, respectively. For boresight rotation, let \(\theta_{z} \) denote the zenith angle (i.e., the angle between the boresight direction and the $z$-axis) and \( \theta_{a} \) denote the azimuth angle (i.e., the angle between the projection of the boresight direction onto the $x$-$y$ plane and the $x$-axis). Accordingly, the components are given by \(f_{x} = \sin\theta_{z} \cos\theta_{a}\), \(f_{y} = \sin\theta_{z} \sin\theta_{a}\), and \(f_{z} = \cos\theta_{z}\). Furthermore, to incorporate practical rotational limitations, the zenith angle of the RA is confined to a specific range \(0 \leq \theta_z \leq \theta_{\max}\), where \(\theta_{\max} \in [0,\frac{\pi}{2}]\) is the maximum adjustable zenith angle.

    \subsection{Channel Model}
	 Given the orientation of RA, the effective antenna gain depends on both the signal arrival/departure direction and the underlying directional gain pattern. We assume that the effective antenna gain of the RA follows a generic directional gain pattern, which is given by\cite{zheng2025rotatableantennaenabledwireless}
     
       \vspace{-0.3cm}
\small	\begin{equation}
		G_e(\epsilon,\varphi)=
		\begin{cases}
			G_{\max}\cos^{2p}(\epsilon), & \epsilon\in[0,\frac{\pi}{2}),\varphi\in[0,2\pi)\\
			0, & \text{otherwise,}  
		\end{cases}
	\end{equation}\normalsize
	where \((\epsilon,\varphi)\) represents a pair of incident angles with respect to the current RA boresight direction. Moreover, \(G_{\max} = 2(2p + 1)\) denotes the maximum gain in the boresight direction obeying the power conservation law, and \( p \) is the directivity factor describing the beamwidth of the antenna’s main lobe. 
    Therefore, the RA's directional gain towards user $k$ is given by \(G_{k} = G_{\max}\left[ \cos(\epsilon_{k})\right]_{+}^{2p}\), where $[x]_+ \triangleq \max\{x,0\}$ denotes the positive operator, with \( \cos(\epsilon_{k}) =	{\vec{\mathbf{f}}}^T \mathbf{\tilde{u}}_{k} \) indicates the inner product of the pointing vector \(\vec{\mathbf{f}}\) and the direction vector \(\mathbf{\tilde{u}}_{k}\).
	
	We assume that the channel from the RA to user $k$ experiences a quasi-static flat-fading, which can be modeled as

       \vspace{-0.2cm}
	\small\begin{equation}
		h_{k}(\boldsymbol{\theta})=\sqrt{\beta_k(r_k) }g_{k}, \label{eq:channel}
	\end{equation}\normalsize
	where \(\boldsymbol{\theta}=[\theta_{z},\theta_{a}]^T\) is the rotational angle vector of the RA, $\beta_k(r_k)$ is the large-scale channel power gain due to the distance-dependent path loss and shadowing, and can be modeled as $\beta_k(r_k)=C_0({r_0}/{r_k})^\alpha$, with $C_0$ being the channel power gain at the reference distance $r_0 = 1$~meter (m), and $\alpha$ being the path loss exponent. Moreover, we assume that $g_k$ is the small-scale channel fading and follows an independent Rician fading distribution, which is characterized by Rician factor $\kappa$ and given by

    \vspace{-0.3cm}
	\small\begin{equation}
    g_k = \sqrt{\frac{\kappa}{\kappa + 1}}\bar{g}_k + \sqrt{\frac{1}{\kappa + 1}}\tilde{g}_k,
    \end{equation}\normalsize
    where $\bar{g}_k = \sqrt{G_{k}}e^{-j\frac{2\pi}{\lambda}r_k}$ denotes the line-of-sight (LoS) component of the channel, and $\tilde{g}_k \sim \mathcal{CN}(0, 1)$ denotes the non-LoS (NLoS) channel component modeled by Rayleigh fading.
	\subsection{NOMA Transmission Scheme}
	In the NOMA scheme, the AP simultaneously transmits signals to both users by adopting superposition coding. The transmitted signal is given by $x = \sqrt{P_1}s_1 + \sqrt{P_2}s_2$, where $P_k$ denotes the power allocated to user $k$, and $s_k$ denotes the data symbol of user $k$ with unit power.
	The received signal at user $k$ can be represented as

 \vspace{-0.3cm}
\small	\begin{equation}
		y_k= h_{k}(\boldsymbol{\theta})(\sqrt{P_1}s_1 + \sqrt{P_2}s_2) + n_k,  \quad k \in \{1,2\}
		\label{eq:yk}
	\end{equation}\normalsize
	where $n_k$ represents the additive white Gaussian noise (AWGN) with zero mean and variance $\sigma^2$. Unlike traditional fixed-antenna systems, the channel gains $h_1(\boldsymbol{\theta})$ and $h_2(\boldsymbol{\theta})$ vary with the RA rotational angle vector $\boldsymbol{\theta}$, leading to $2!=2$ possible permutations for the user decoding order. Given the target rates of the two users, we aim to minimize the total transmit power at the AP by optimizing the rotational angle vector $\boldsymbol{\theta}$, subject to the rotational constraints. 
	
	Thus, the corresponding optimization problem can be decomposed into two subproblems associated with two different decoding orders as follows.
    
    \vspace{-0.3cm}
	\small
	\begin{subequations}
		\begin{align}
			(\text{N1}): P_{N1} \triangleq\min_{P_1, P_2, \boldsymbol{\theta}}  & P_1 + P_2 \\
			\text{s.t. } \quad & \log_2\left(1+\frac{P_1 |h_1(\boldsymbol{\theta})|^2}{P_2 |h_1(\boldsymbol{\theta})|^2 + \sigma^2}\right) \ge \gamma_1 \\
			& \log_2\left(1+\frac{P_2 |h_2(\boldsymbol{\theta})|^2}{\sigma^2}\right) \ge \gamma_2\\
			&\ 0 \leq \theta_{z} \leq \theta_{\max} \label{eq:con_angle1}\\
			&\ 0 \leq \theta_{a} \leq 2\pi \label{eq:con_angle2}
		\end{align}
	\end{subequations}\normalsize 	\vspace{-0.35cm}
\small	\begin{subequations}
	\begin{align}
		(\text{N2}): P_{N2} \triangleq\min_{P_1, P_2, \boldsymbol{\theta}}  & P_1 + P_2 \\
		\text{s.t.}  \quad & \log_2\left(1+\frac{P_2 |h_2(\boldsymbol{\theta})|^2}{P_1 |h_2(\boldsymbol{\theta})|^2 + \sigma^2}\right) \ge \gamma_2 \\
		& \log_2\left(1+\frac{P_1 |h_1(\boldsymbol{\theta})|^2}{\sigma^2}\right) \ge \gamma_1\\
		& \eqref{eq:con_angle1}, \eqref{eq:con_angle2} \notag
	\end{align}
\end{subequations}\normalsize
where $\gamma_1$ and $\gamma_2$ are the target rates of users 1 and 2 in bits per second per Hertz (bps/Hz), respectively. Since the user rates are monotonically increasing with $P_1$ and $P_2$, the inequality rate constraints should be met with equality at the optimal solution. By eliminating the equality constraints, (\text{N1}) and (\text{N2}) can be simplified as

\vspace{-0.3cm}
\small	
	\begin{align}
		(\text{N1.1}):	P_{N1} \triangleq\min_{\boldsymbol{\theta}} & \frac{(2^{\gamma_1}-1) \sigma^2}{|h_1(\boldsymbol{\theta})|^2} + \frac{(2^{\gamma_2}-1) 2^{\gamma_1} \sigma^2}{|h_2(\boldsymbol{\theta})|^2} \label{eq:N1.1}\\
		\text{s.t.} \quad&\eqref{eq:con_angle1}, \eqref{eq:con_angle2} \notag
	\end{align}

\normalsize 	\vspace{-0.35cm}
\small	
	\begin{align}
		(\text{N2.1}):	P_{N2} \triangleq\min_{\boldsymbol{\theta}} & \frac{(2^{\gamma_2}-1) \sigma^2}{|h_2(\boldsymbol{\theta})|^2} + \frac{(2^{\gamma_1}-1) 2^{\gamma_2} \sigma^2}{|h_1(\boldsymbol{\theta})|^2} \label{eq:N2.1}\\
		\text{s.t.} \quad&\eqref{eq:con_angle1}, \eqref{eq:con_angle2} \notag
	\end{align}

	\normalsize
	Finally, the minimum transmit power required by NOMA is obtained as
\small	\begin{equation}
		P_N \triangleq \min \{ P_{N1}, P_{N2} \}.\label{eq:Nclosed}
	\end{equation}
	\normalsize
	
	\subsection{OMA Transmission Schemes}
	\subsubsection{FDMA}
	As shown in Fig. 1, the AP serves the two users simultaneously over two equal frequency-domain adjacent RBs via FDMA. Accordingly, the optimization problem of minimizing the total transmit power at the AP is formulated as

    \vspace{-0.3cm}
	\small\begin{subequations}
		\begin{align}
			(\text{F1}): P_{F} \triangleq \min_{P_1, P_2, \boldsymbol{\theta}} \quad & P_1 + P_2 \\
			\text{s.t. } \quad & \frac{1}{2} \log_2\left(1 + \frac{P_k |h_k(\boldsymbol{\theta})|^2}{\frac{1}{2} \sigma^2}\right) \ge \gamma_k, \forall k \label{eq:F1}\\
			&\eqref{eq:con_angle1}, \eqref{eq:con_angle2}\notag
		\end{align}
	\end{subequations}\normalsize
	Note that the factor 1/2 in \eqref{eq:F1} is due to the fact that each user is assigned with half of the bandwidth as compared to NOMA. Similar to the NOMA scheme, the relevant inequality rate constraint can be transformed into equality constraints. Accordingly, the power minimization problem for FDMA can be formulated as

    \vspace{-0.3cm}
	\small
	\begin{align}
	(\text{F1.1}):	P_{F} \triangleq \min_{\boldsymbol{\theta}} & \frac{\sigma^2\left(2^{2\gamma_1}-1\right)}{2 |h_1(\boldsymbol{\theta})|^2}+ \frac{\sigma^2\left(2^{2\gamma_2}-1\right)}{2 |h_2(\boldsymbol{\theta})|^2} \label{eq:F1.1}\\
	\text{s.t.}	\quad&\eqref{eq:con_angle1}, \eqref{eq:con_angle2}\notag
		\label{eq:PFtheta}
		\end{align}
	\normalsize

	\subsubsection{TDMA}
	As shown in Fig. 1, the AP communicates with two users consecutively over two equal time-domain adjacent RBs via TDMA. The key difference of integrating RA with TDMA is the time-switching capability, which allows the RA to dynamically adjust its boresight direction for each user in their respective time slots. Thus, unlike the NOMA and FDMA schemes, where an identical rotational angle vector $\boldsymbol{\theta}$ is applied to both users, TDMA involves two optimization variables $\boldsymbol{\theta}_1$ and $\boldsymbol{\theta}_2$. Therefore, the optimization problem of minimizing the total transmit power at the AP in the case of TDMA can be formulated as

    \vspace{-0.3cm}
	\small	\begin{subequations}
		\begin{align}
			(\text{T1}): P_{T} \triangleq \min_{P_1, P_2, \boldsymbol{\theta}_1, \boldsymbol{\theta}_2} & P_1 + P_2 \\
			\text{s.t.} \quad & \frac{1}{2} \log_2\left(1 + \frac{2P_k |h_k(\boldsymbol{\theta}_{k})|^2}{ \sigma^2}\right) \ge \gamma_k, \forall k\label{eq:T1}\\
			&\eqref{eq:con_angle1}, \eqref{eq:con_angle2}\notag
		\end{align}
	\end{subequations}	\normalsize
Note that the factor 1/2 in \eqref{eq:T1} is due to the fact that each user is assigned half of the time as compared to NOMA. Similar to (\text{N1}), (\text{T1}) can be transformed into

\vspace{-0.3cm}
	\small
	\begin{align}
		(\text{T1.1}): P_{T} \triangleq	\min_{\boldsymbol{\theta}_1} & \frac{\sigma^2\left(2^{2\gamma_1}-1\right)}{2 |h_1(\boldsymbol{\theta}_1)|^2} + \min_{\boldsymbol{\theta}_2}\frac{\sigma^2\left(2^{2\gamma_2}-1\right)}{2 |h_2(\boldsymbol{\theta}_2)|^2} \label{eq:T1.1}\\
	\text{s.t.}	\quad&\eqref{eq:con_angle1}, \eqref{eq:con_angle2}\notag
		\label{eq:PT}
		\end{align}
\normalsize

	\section{Comparison of Minimum Transmit Powers}
	\label{sec:comparison}
	\subsection{TDMA vs. FDMA}
	First, we compare the minimum transmit powers required by RA-assisted FDMA and TDMA, whose relationship is given by the following proposition.
	
	\textbf{Proposition 1:} The minimum transmit power required by FDMA is always no less than that by TDMA, i.e., $P_F \ge P_T$. 

	\textit{Proof:}  Let $P_k(\boldsymbol{\theta}) = \frac{(2^{2\gamma_k}-1)\sigma^2}{2|h_k(\boldsymbol{\theta})|^2}$ be the power consumption component for user $k$. Then, the FDMA power consumption in \eqref{eq:F1.1} can be written as $P_F = \min_{\boldsymbol{{\theta}}_F} (P_1(\boldsymbol{\theta}_{F}) + P_2(\boldsymbol{\theta}_{F}))$. In contrast, the TDMA power consumption in \eqref{eq:T1.1} allows for individual optimization, and can be written as $P_T = \min_{\boldsymbol{\theta}_1} P_1(\boldsymbol{\theta}_1) + \min_{\boldsymbol{\theta}_2} P_2(\boldsymbol{\theta}_2)$. Since the minimum of the sum of two functions is always no less than the sum of their individual minimum values (i.e., $\min(A+B) \ge \min A + \min B$), it follows that $P_F \ge P_T$. The equality holds if and only if the optimal rotational angles are identical, i.e., $\boldsymbol{\theta}^*_{1} = \boldsymbol{\theta}^*_{2} = \boldsymbol{\theta}^*_F$. \hfill $\blacksquare$
	
	\subsection{NOMA vs. FDMA}
	Next, we compare the minimum transmit powers required by NOMA and FDMA for the RA-assisted system, with the results given by the following proposition.
	
	\textbf{Proposition 2:} The minimum transmit power required by NOMA is no more than that by FDMA, i.e., $P_F\geq P_N$, and the equality holds if and only if \(\boldsymbol{\theta}_N = \boldsymbol{\theta}_F=\boldsymbol{\theta}^*\), $|h_1(\boldsymbol{\theta}^*)| = |h_2(\boldsymbol{\theta}^*)|$, and \(\gamma_1=\gamma_2\), where \(\boldsymbol{\theta}_N\) and \(\boldsymbol{\theta}_F\) represent the optimal solutions to problem  (\text{N1.1}) and  (\text{F1.1}), respectively.
	
	\textit{Proof:}  First, under the channel condition of $|h_1(\boldsymbol{\theta}^*)| \le |h_2(\boldsymbol{\theta}^*)|$, we consider the power gap between \eqref{eq:N1.1} and \eqref{eq:F1.1}, which is given by

    \vspace{-0.3cm}
	\small	\begin{align}
		\Delta P_1 &= P_{F}-P_{N1}\notag\\
		&\overset{(a)}{\operatorname*{\operatorname*{\geq}}} \frac{(2^{2\gamma_1}-1)\sigma^2}{2 |h_1(\boldsymbol{\theta}^*)|^2} + \frac{(2^{2\gamma_2}-1)\sigma^2}{2 |h_2(\boldsymbol{\theta}^*)|^2} 
		 -  \frac{(2^{\gamma_1}-1)\sigma^2}{|h_1(\boldsymbol{\theta}^*)|^2}\notag\\ &\quad+ \frac{(2^{\gamma_2}-1)2^{\gamma_1}\sigma^2}{|h_2(\boldsymbol{\theta}^*)|^2}\notag\\
		&= \frac{\sigma^2\left[ (2^{2\gamma_2}-1) - 2(2^{\gamma_2}-1)2^{\gamma_1} \right]}{2 |h_2(\boldsymbol{\theta}^*)|^2} + \frac{\sigma^2(2^{\gamma_1}-1)^2}{2 |h_1(\boldsymbol{\theta}^*)|^2}  \notag\\
		&\overset{(b)}{\operatorname*{\operatorname*{\geq}}} \frac{\sigma^2}{2 |h_2(\boldsymbol{\theta}^*)|^2} \left[ (2^{\gamma_1}-1)^2 + 2^{2\gamma_2}-1 - 2^{\gamma_2+\gamma_1+1} + 2^{\gamma_1+1} \right] \nonumber \\
		&= \frac{\sigma^2}{2 |h_2(\boldsymbol{\theta}^*)|^2} (2^{\gamma_1} - 2^{\gamma_2})^2 \overset{(c)}{\operatorname*{\operatorname*{\geq}}} 0
	\end{align}\normalsize
where the equality of ($a$) holds if the optimal solution to problem (\text{N1.1}) is equal to the optimal solution to problem (\text{F1.1}), which yields \(\boldsymbol{\theta}_N = \boldsymbol{\theta}_F=\boldsymbol{\theta}^*\); the equality of ($b$) holds when $|h_2(\boldsymbol{\theta}^*)|= |h_1(\boldsymbol{\theta}^*)|$; and the equality of ($c$) holds when \(\gamma_1=\gamma_2\). Similarly, for the channel condition of $|h_1(\boldsymbol{\theta}^*)| \ge |h_2(\boldsymbol{\theta}^*)|$, we can also obtain a non-negative power gap between \eqref{eq:N2.1} and \eqref{eq:F1.1} following the same method, i.e., $\Delta P_2 = P_F - P_{N2}\geq 0$. Since $P_F - P_{N1}\geq 0$ and $P_F - P_{N2}\geq 0$, we can conclude that $P_F\geq \min\{P_{N1},P_{N2}\}=P_{N}$, where the equality holds when all the above conditions are satisfied.	\hfill $\blacksquare$

\section{Proposed PSO-Based Solution}

In \Cref{sec:comparison}, we have derived the closed-form expressions for the minimum transmit power for any given rotational angles (i.e., \eqref{eq:Nclosed}, \eqref{eq:F1.1}, and \eqref{eq:T1.1}), the remaining power minimization problems with respect to $\boldsymbol{\theta}$ are still non-convex. To address this, we employ a PSO-based algorithm to optimize the rotational angle vector in this section.

\subsection{Particle Representation and Fitness Function}

In the employed PSO framework, a swarm of $S$ particles is initialized \cite{PSO}. Each particle represents a candidate solution for the RA's rotational angle configuration. The structure of the search space varies by the multiple access scheme:

\subsubsection{For NOMA and FDMA}
Since the two schemes utilize a common rotational angle vector $\boldsymbol{\theta}$ for both users, the position vector of particle $i$ at the $t$-th iteration can be expressed as

\vspace{-0.3cm}
	\small\begin{equation}
	\boldsymbol{x}_i^{(t)} = [\theta_{z,i}^{(t)}, \theta_{a,i}^{(t)}]^T.
\end{equation}\normalsize
The fitness function is defined as the minimum transmit power derived in \Cref{sec:system}, and can be represented as

\vspace{-0.3cm}
\small\begin{equation}
	Y(\boldsymbol{x}_i^{(t)}) = P_{\chi}(\boldsymbol{\theta}_i^{(t)}), \quad \chi \in \{N, F\},
\end{equation}\normalsize
where $P_N$ is calculated based on the closed form in \eqref{eq:Nclosed}, taking the minimum of the two decoding orders, and $P_F$ is calculated by using \eqref{eq:F1.1}.

\subsubsection{For TDMA}
Leveraging the time-switching capability, the RA can adjust its rotational angles independently for each user's time slot. Thus, the corresponding particle position vector is defined as

\vspace{-0.3cm}
\small\begin{equation}
	\boldsymbol{x}_i^{(t)} = [\theta_{z_1,i}^{(t)}, \theta_{a_1,i}^{(t)}, \theta_{z_2,i}^{(t)}, \theta_{a_2,i}^{(t)}]^T.
\end{equation}\normalsize
The fitness function for the total TDMA power $P_T$ is 
given in \eqref{eq:T1.1}.

\subsection{Updating Process}

The algorithm updates the velocity $\mathbf{v}_i$ and position $\mathbf{x}_i$ of each particle iteratively to converge towards the global optimum. The update rules are given by

\vspace{-0.3cm}
\small\begin{align}
	\boldsymbol{v}_i^{(t+1)} &= w \boldsymbol{v}_i^{(t)} + c_1 r_1 (\boldsymbol{p}_{\mathrm{best},i} - \boldsymbol{x}_i^{(t)}) + c_2 r_2 (\boldsymbol{g}_{\mathrm{best}} - \boldsymbol{x}_i^{(t)}), \label{eq:v}\\ 
	\boldsymbol{x}_i^{(t+1)} &= \boldsymbol{x}_i^{(t)} + \boldsymbol{v}_i^{(t+1)}, \label{eq:x}
\end{align}\normalsize
where $w$ represents the inertia weight, $c_1$ and $c_2$ are the cognitive and social acceleration coefficients, respectively, and $r_1, r_2 \sim U(0,1)$ are random scaling factors. Moreover, $\boldsymbol{p}_{\mathrm{best},i}$ denotes the personal best position achieved by particle $i$ so far, while $\boldsymbol{g}_{\mathrm{best}}$ denotes the global best position found by the entire swarm. 

The iterative updating process continues until a termination criterion is achieved, typically when a maximum number of iterations $T_{\max}$ is reached or the improvement in the fitness value becomes negligible. Upon termination, the final global best position $\boldsymbol{g}_{\mathrm{best}}$ is selected as the optimized antenna rotational angle $\boldsymbol{\theta}^\star$, and its corresponding fitness value represents the achieved global minimum transmit power $P^\star$. \footnote{Due to the heuristic nature of PSO, the obtained $\boldsymbol{g}_{\mathrm{best}}$ is a near-optimal solution, whose gap to the theoretical global optimum diminishes with increased swarm size $S$ and iterations $T_{\max}$\cite{PSO}.}

\subsection{Overall Algorithm}

\begin{algorithm}
	\small
	\caption{PSO Algorithm for Rotational Angle Optimization}
	\label{alg:pso}
	\begin{algorithmic}[1]
		\State \textbf{Input:} $S, T_{\max}, w, c_1, c_2$.
		
		\State \textbf{Initialize:} $\boldsymbol{x}_i^{(0)}, \boldsymbol{v}_i^{(0)}$; $\boldsymbol{p}_{\mathrm{best},i} \leftarrow \boldsymbol{x}_i^{(0)}$; $\boldsymbol{g}_{\mathrm{best}}$.
		
		\For{$t = 1$ to $T_{\max}$}
		\For{$i = 1$ to $S$}
		\State Update $\boldsymbol{v}_i^{(t)}$ and $\boldsymbol{x}_i^{(t)}$ via rules in \eqref{eq:v} and \eqref{eq:x}.
		\If{$Y(\boldsymbol{x}_i^{(t)}) < Y(\boldsymbol{p}_{\mathrm{best},i})$} $\boldsymbol{p}_{\mathrm{best},i} \leftarrow \boldsymbol{x}_i^{(t)}$; \EndIf
		\If{$Y(\boldsymbol{p}_{\mathrm{best},i}) < Y(\boldsymbol{g}_{\mathrm{best}})$} $\boldsymbol{g}_{\mathrm{best}} \leftarrow \boldsymbol{p}_{\mathrm{best},i}$; \EndIf
		\EndFor
		\EndFor
		\State \textbf{Return} Optimal angle $\boldsymbol{\theta}^\star \leftarrow \boldsymbol{g}_{\mathrm{best}}$.
	\end{algorithmic}
\end{algorithm}
 The detailed procedures are summarized in Algorithm \ref{alg:pso}. The computational complexity of the employed PSO-based algorithm depends on the swarm size $S$, the maximum number of iterations $T_{\max}$, and the search space dimension $D$. For the RA-assisted NOMA and FDMA schemes, the search dimension is $D=2$ (corresponding to $\theta_z$ and $\theta_a$), whereas for the TDMA scheme, the dimension increases to $D=4$ due to the independent rotational angle vector for each time slot. Consequently, the overall complexity is $\mathcal{O}(T_{\max}SD)$.

\begin{figure*}[t!]
		\centering
		\begin{minipage}[t]{0.49\textwidth}
			\centering
			\begin{subfigure}[b]{0.49\linewidth}
				\centering
				\includegraphics[width=\linewidth]{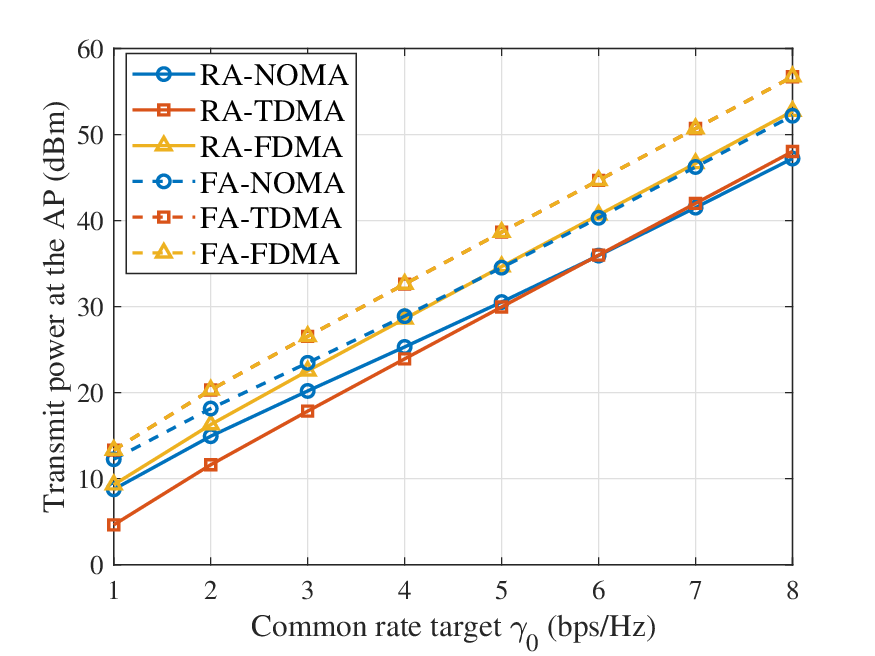} 
				\caption{AP transmit power versus the common target rate $\gamma_0$.\\	}
				\label{Fig:2(a)}
			\end{subfigure}
			\hfill
			\begin{subfigure}[b]{0.49\linewidth}
				\centering
				\includegraphics[width=\linewidth]{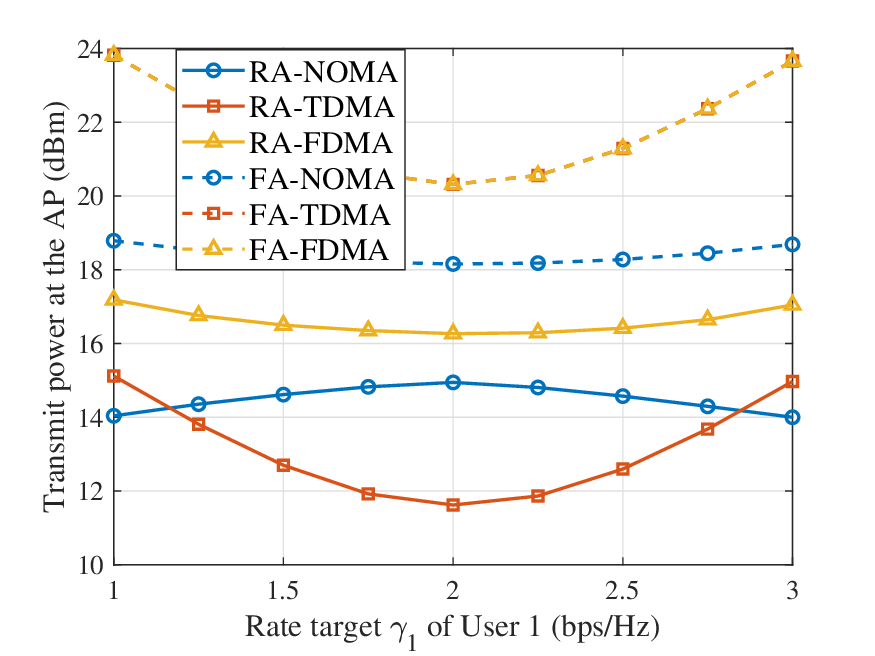}
				\caption{AP transmit power versus user 1's target rate $\gamma_1$, with the two users' sum rate fixed as $\gamma_1 + \gamma_2 = 4$ bps/Hz.}
				\label{fig:2(b)}
			\end{subfigure}
			\captionsetup{labelformat=empty, justification=raggedright, singlelinecheck=false}
             \vspace{-0.1cm}
			\caption{Fig. 2. Performance comparison of OMA and NOMA in Case 1.}
			\label{fig:case1_all}
		\end{minipage}
             \vspace{-0.25cm}
		\hfill 
		\begin{minipage}[t]{0.49\textwidth}
			\centering
			\begin{subfigure}[b]{0.49\linewidth}
				\centering
				\includegraphics[width=\linewidth]{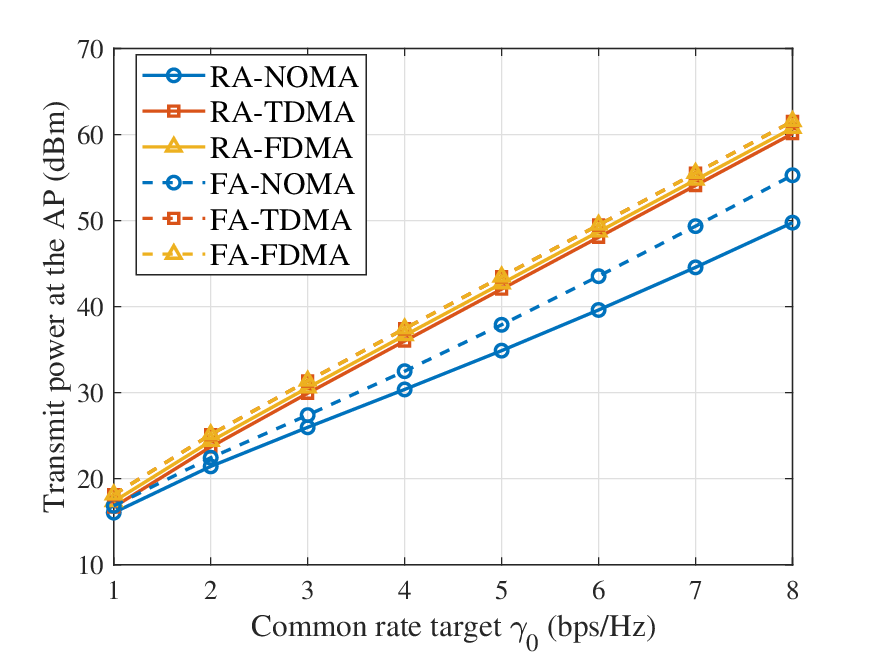}
				\caption{AP transmit power versus the common target rate $\gamma_0$.\\}
				\label{fig:3(a)}
			\end{subfigure}
			\hfill
			\begin{subfigure}[b]{0.49\linewidth}
				\centering
				\includegraphics[width=\linewidth]{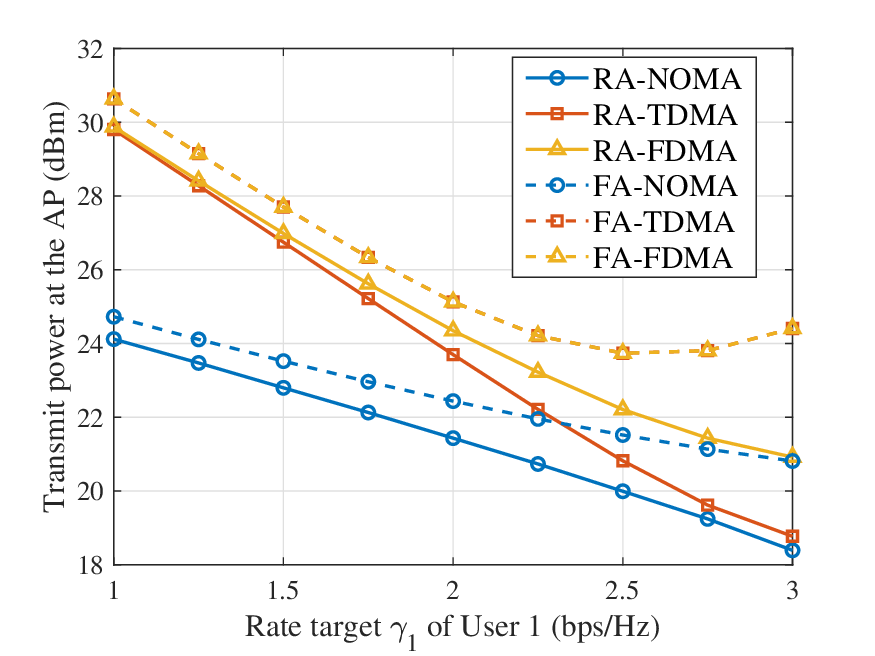}
				\caption{AP transmit power versus user 1's target rate $\gamma_1$, with the two users' sum rate fixed as $\gamma_1 + \gamma_2 = 4$ bps/Hz.}
				\label{fig:3(b)}
			\end{subfigure}
				\captionsetup{labelformat=empty, justification=raggedright, singlelinecheck=false}
                \vspace{-0.1cm}
			\caption{Fig. 3. Performance comparison of OMA and NOMA in Case 2.}
			\label{fig:case2_all}
		\end{minipage}
             \vspace{-0.25cm}
	\end{figure*}

	\section{Simulation Results}

	In this section, numerical results are presented to validate our analytical results and compare the proposed RA-assisted NOMA (i.e., RA-NOMA) with its OMA counterparts, namely RA-TDMA and RA-FDMA. Additionally, we compare these schemes with conventional fixed-antenna baselines, namely FA-NOMA, FA-TDMA, and FA-FDMA, to quantify the performance gain brought by antenna rotation. Unless otherwise stated, we set the simulation parameters as follows: the wavelength is $\lambda = 0.125$ m, the noise power is ${\sigma}^2=-80 $~dBm, the antenna directivity factor is $p = 2$, the path loss exponent is $\alpha = 3.2$, the reference channel power gain is $C_0 = -40$ dB, and the Rician factor is set as $\kappa=1$.
	The simulation results are categorized into two deployment scenarios: 
	1) \textbf{Symmetric Case 1}: Two users are symmetrically located at $\mathbf{u}_1 = [20, 20, 0]^T$~m and $\mathbf{u}_2 = [20, -20, 0]^T$~m, having identical distances to the AP.
	2) \textbf{Asymmetric Case 2}: Two users are located at $\mathbf{u}_1 = [20, 20, 0]^T$~m and $\mathbf{u}_2 = [80, -20, 0]^T$~m, representing a typical near-far scenario with significant channel disparity. The performance results below are averaged over 1000 independent random realizations to eliminate fluctuations.

	\cref{Fig:2(a)} depicts the total transmit power versus the common target rate $\gamma_0$ (i.e., $\gamma_1 = \gamma_2 = \gamma_0$). It is observed that all RA-assisted schemes significantly outperform their FA counterparts. This is because RA enables the AP to flexibly adjust the antenna boresight toward favorable propagation directions, thereby maximizing the effective channel gain and effectively reducing the transmit power consumption. Moreover, among RA-assisted schemes, NOMA and TDMA require lower transmit power, while FDMA exhibits the highest transmit power. This is consistent with the theoretical results in \Cref{sec:comparison}, i.e., $P_N^\star \le P_F^\star$ and $P_T^\star \le P_F^\star$. However, it is worth noting that in this symmetric scenario, the required transmit power by RA-TDMA is not always higher than that of RA-NOMA. This is because RA enables boresight adjustment across TDMA time slots, allowing sequential alignment with each user and thereby partially compensating for the spectral efficiency loss compared to NOMA.
	
	To draw more insight into Case 1, \cref{fig:2(b)} shows the required transmit power at the AP versus user 1's target rate $\gamma_1$, with the sum rate fixed at $\gamma_1 + \gamma_2 = 4$ bps/Hz. First, it is observed that the RA-assisted schemes demonstrate a consistent advantage over their FA counterparts. Moreover, we can observe that TDMA requires lower transmit power than NOMA when the target rates of the two users are symmetric (i.e., near $\gamma_1=2$ bps/Hz). Furthermore, the results show that the RA-assisted NOMA scheme is highly robust to rate asymmetry, maintaining a low and stable transmit power across the entire range. In contrast, the power consumption of RA-assisted TDMA and FDMA increases sharply as the rate pairs become more asymmetric (i.e., moving away from $\gamma_1=2$ bps/Hz). This is due to the fact that, by allowing the user with higher rate demand to utilize the full bandwidth, NOMA can efficiently accommodate asymmetric transmission demands, whereas OMA schemes suffer from inefficient resource allocation in such cases.
	
	In contrast to the symmetric deployment in Case 1, we consider an asymmetric scenario in Case 2 with one near-user and one far-user. Due to the severe path loss experienced by the far user, the total transmit power required at the AP increases drastically, as shown in \cref{fig:case2_all}, compared to the symmetric case in \cref{fig:case1_all}. More importantly, the results show that NOMA consistently requires lower transmit power than OMA schemes for both TDMA and FDMA. This superiority arises because NOMA can exploit the pronounced channel disparity in asymmetric environments, enabling efficient SIC and thereby reducing the required transmit power.
	
	\section{Conclusion}
	In this letter, we investigated the transmit power minimization problems for both NOMA and OMA schemes in an RA-assisted two-user communication system. We analytically compared the minimum transmit powers required by different multiple access schemes and employed a PSO-based algorithm to optimize the rotational angles. Simulation results demonstrated that RA can significantly reduce the transmit power compared to conventional fixed-antenna benchmarks. Moreover, our analysis revealed that NOMA always has superior performance to FDMA, while TDMA may outperform NOMA for symmetrically deployed users with common target rates by employing the time-switching feature of RA. 
	
	\bibliographystyle{IEEEtran} 
    \bibliography{references} 
\end{document}